\documentclass[12pt,preprint]{aastex}
\usepackage{natbib}
\usepackage{epsf}

\begin{document}
\title{\Large\bf High Redshift Radio Galaxies: Laboratories for 
Massive Galaxy and Cluster Formation in the early Universe}
\vskip 0.2in

\noindent{G. Miley (Leiden), 
C. Carilli\footnote{Contact author: Chris Carilli, National Radio
Astronomy Observatory, P.O.Box O, Socorro, NM, 87801,
ccarilli@nrao.edu} (NRAO),  G.B. Taylor (UNM), 
C. de Breuck (ESO), \& \\ A. Cohen (NRL)}
\vskip 5.2in

\includegraphics{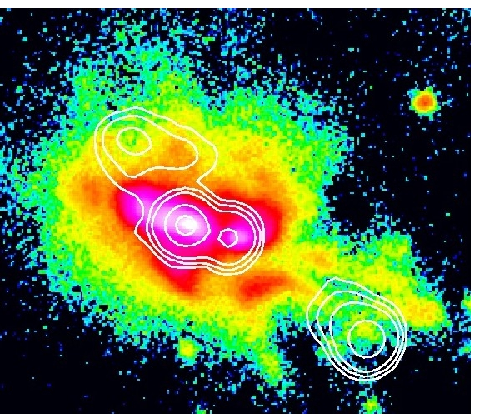}
\includegraphics{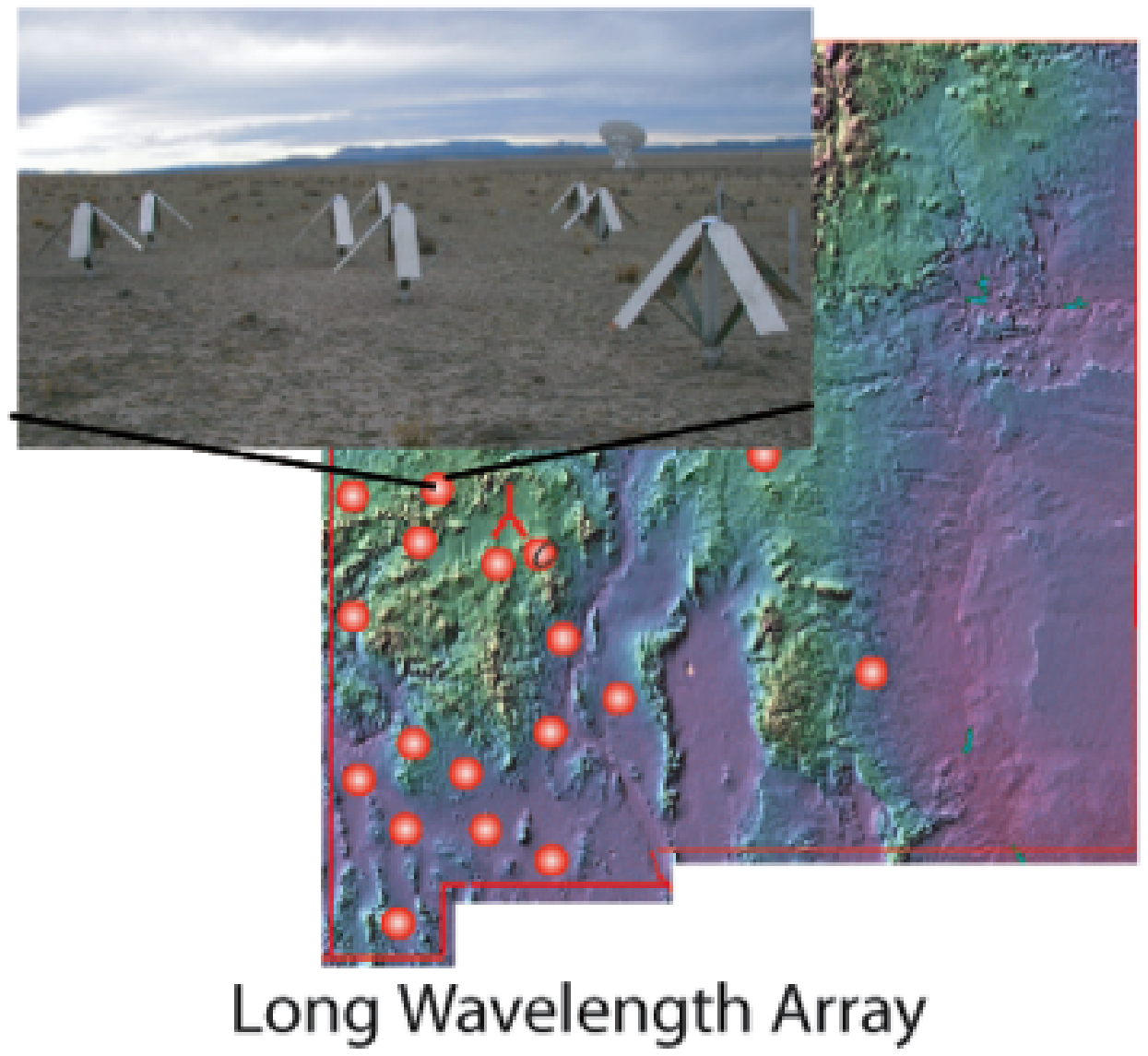}
\includegraphics{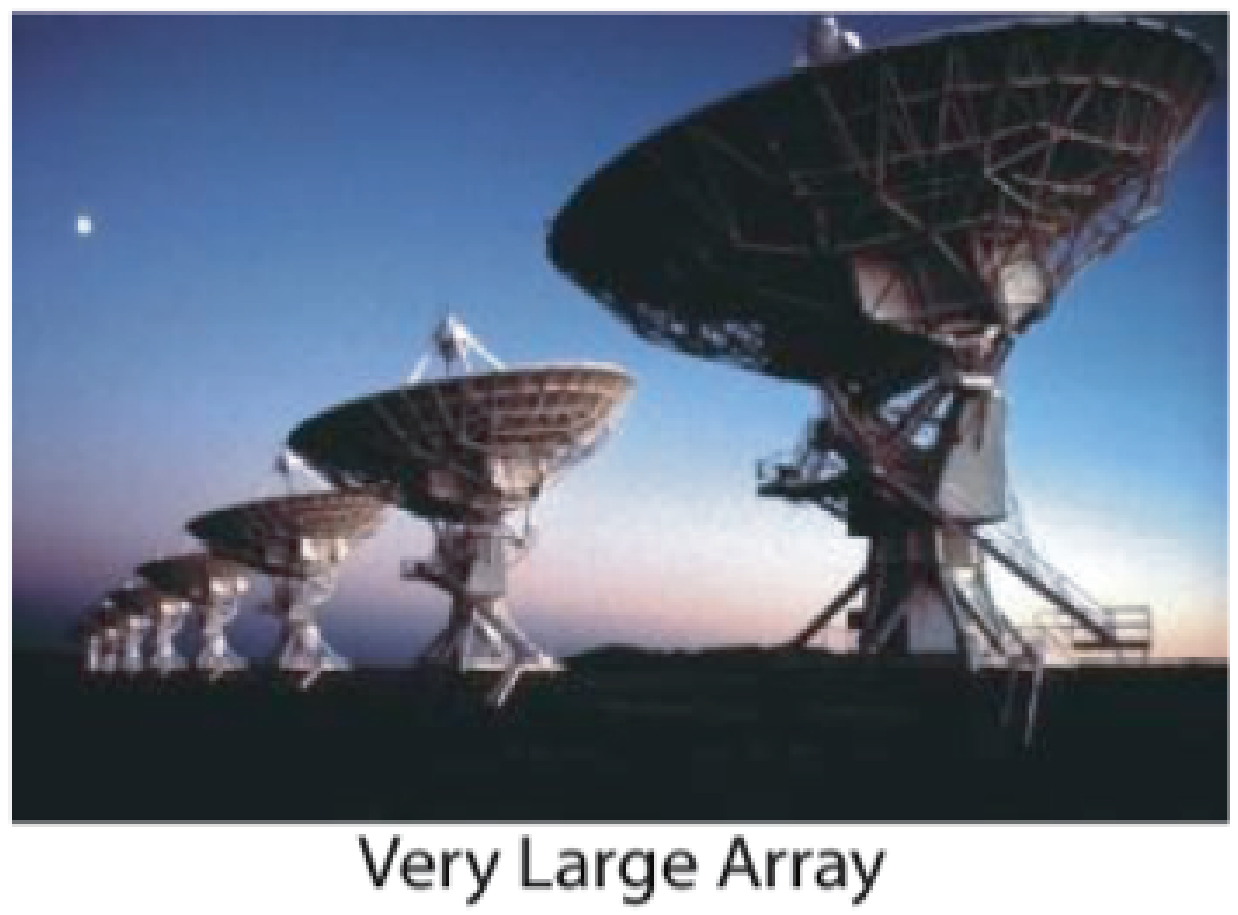}
\includegraphics{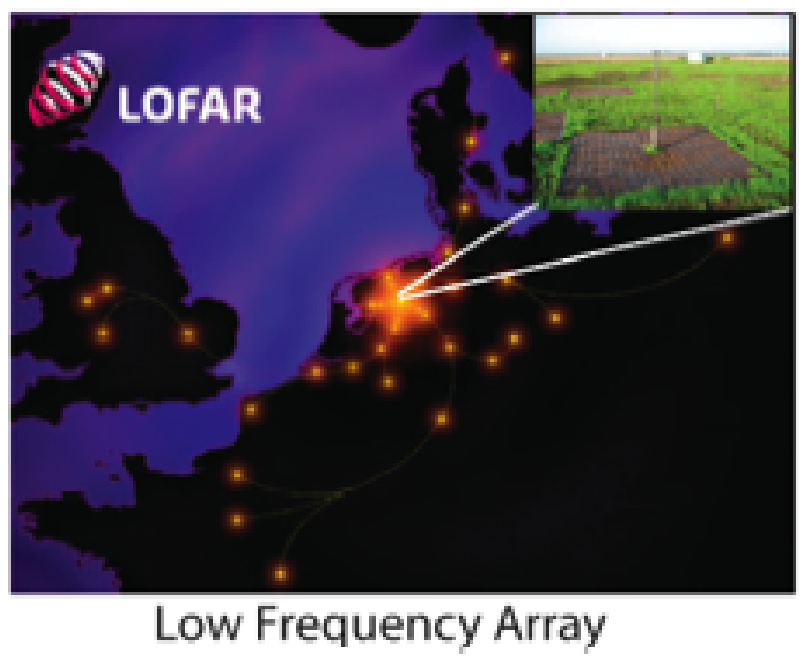}
\includegraphics{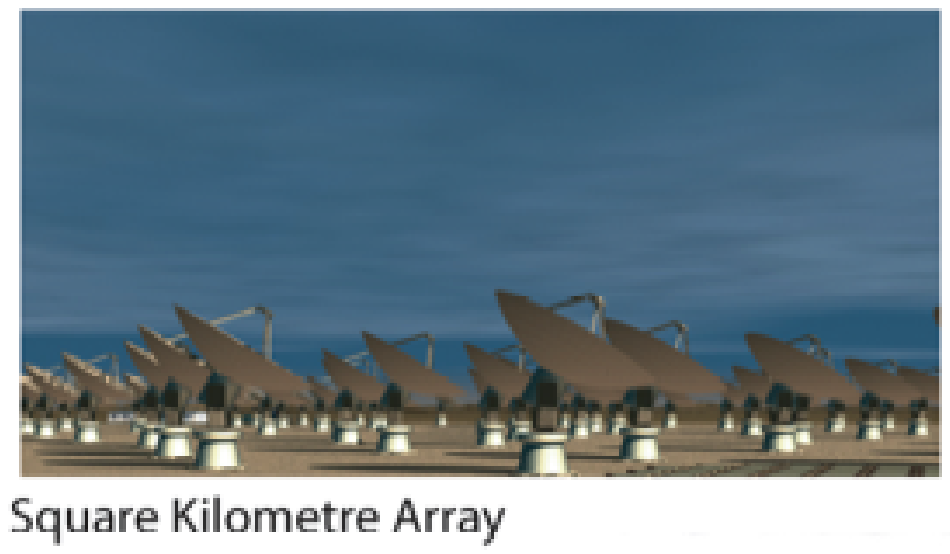}
\includegraphics{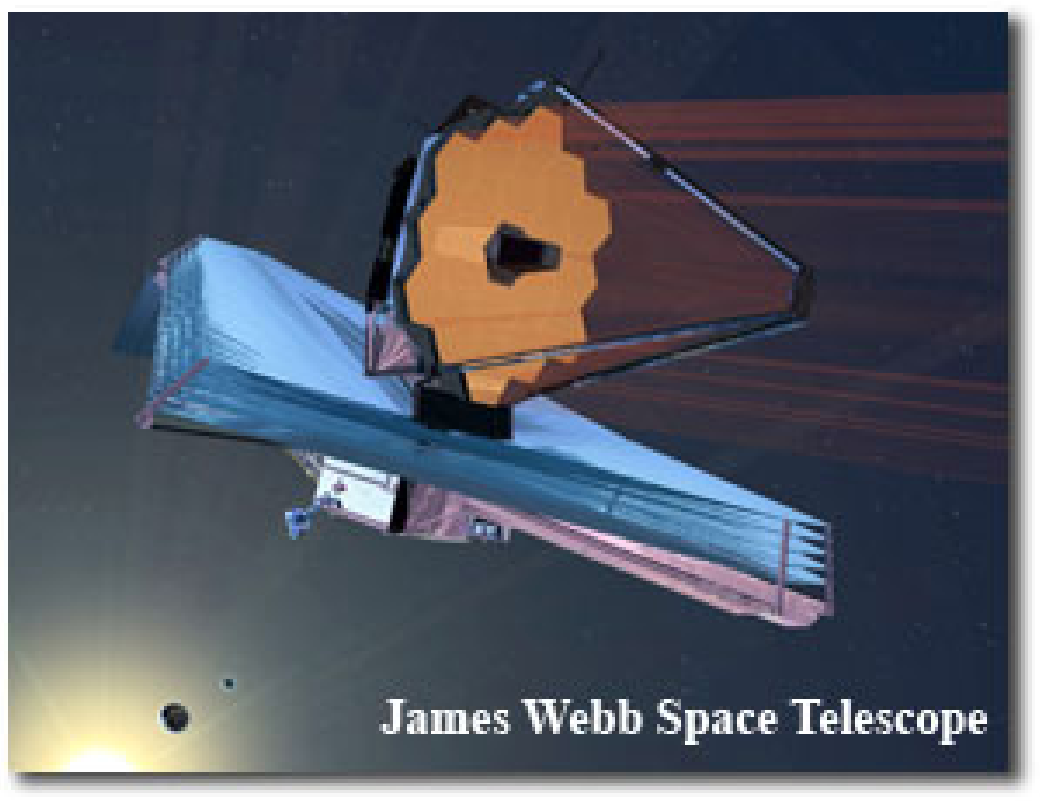}

\clearpage
\noindent

\centerline{\bf 1. Beacons to massive galaxy and
cluster formation in the early Universe}

High redshift radio galaxies (HzRGs, z $>$ 2) are among the largest, most luminous, 
most massive, and most beautiful objects in the Universe.  They are 
generally identified from their radio emission, thought to be powered by 
accretion of matter onto supermassive black holes in the nuclei of their 
host galaxies.  Further observations show that they are energetic
sources of radiation throughout most of the electromagnetic spectrum.  
Figure 1 shows the spectral energy distribution (SED) of a typical
HzRG from radio to X-ray wavelengths, together with a decomposition
into various observable HzRG constituents - relativistic plasma, gas
and dust, stars and the active galactic nuclei (AGN).  

In a recent review article Miley \& de Breuck (9) present an extensive
analysis of the properties and implications of HzRGs, including a
detailed description of the major emitting components which provide
important diagnostics about various physical constituents of the early
Universe.  A list of known HzRG building blocks is given in Table 1,
together with a summary of techniques used to study them. Also
included are a list of the resultant diagnostics, some useful
references (see Miley \& De Breuck), and our best estimate for the
typical mass of the component in HzRGs.  As can be seen in Table 1,
several constituents of HzRGs are inferred to be extremely massive,
including old stars (up to $\sim$ 10$^{12}$ M$_{\odot}$), hot gas (up
to $\sim$ 10$^{12}$ M$_{\odot}$) and molecular gas (up to $\sim$
10$^{11}$ M$_{\odot}$).

\begin{table}[htbp]
\label{table: constituents}
\begin{center}
{\bf{Table 1. Constituents of distant radio galaxies}}
\label{diagtable}
\\ [3ex]
\begin{tabular}{|p{90pt}|p{90pt}|p{165pt}|p{30pt}|p{33pt}|}
\hline
\textbf{Constituent}&
\textbf{Observable}&
\textbf{Typical Diagnostics}&
\textbf{Refs.}&
\textbf{Mass}
\par (M$_\odot$)
\\
[1ex]
\hline
\hline
Relativistic plasma&
Radio continuum&
Magnetic field, age, energetics, pressure, particle acceleration. Jet collimation and propagation&
1,2 &
\\
\cline{2-4}
 &
X-ray continuum&
Magnetic field, equipartition, pressures&
3,4,1 &
  \\
\hline
\hline
Hot ionized gas \par T$_{e}$ $\sim$ $10^{7}$--10$^{8}$K \par n$_{e}$ $\sim$ 10$^{-1.5}$cm$^{-3}$&
Radio (de)polarisation&
Density, magnetic field, &
1 &
$^{}$ \par 10$^{11 - 12}$
\\
\cline{2-4}
 &
X-rays&
Temperature, density mass&
&
  \\
\hline
Warm ionized gas \par T$_{e}$ $\sim$ 10$^{4}$--10$^{5}$K \par n$_{e}$ $\sim $ 10$^{0.5 - 1.5}$cm$^{-3}$&
UV-optical \par emission lines \par &
Temperature, density, kinematics, mass, ionisation, metallicity, filling factor&
5,6,7,8&
$^{}$ \par 10$^{9 - 10.5}$
 \\
\cline{2-4}
&
Nebular \par continuum&
SED contamination&
9,10&
\\
\hline
Cool atomic gas \par T$_{s}$ $\sim$10$^{3}$K \par n(HI) $\sim$ 10$^{1}$cm$^{-3}$&
HI absorption&
Kinematics, column densities, spin temperature, sizes, mass&
11,8&
$^{}$ \par 10$^{7 - 8}$
\\
\cline{2-4}
&
UV-optical \par absorption lines&
Kinematics, mass, column densities, metallicity&
8,12  13,14&
\\
\hline
Molecular gas \par T $\sim$ 50 - 500K \par
n(H$_{2}$) $>$ 10$^{2}$ cm$^{-3}$&
(Sub)millimeter lines&
Temperature, density, mass&
15
&
$^{}$ \par 10$^{10-11}$
 \\
\hline
\hline
Dust \par
T $\sim$ 50 - 500K &
UV-optical \par polarisation&
Dust composition, scattering, mass, hidden quasar&
16  17&
$^{}$ \par 10$^{8 - 9}$
\par
%Check
%\raisebox{-1.50ex}[0cm][0cm]{}
\\
\cline{2-4}
 &
(Sub)millimeter continuum&
Temperature, mass, heating source&
18&
  \\
\hline
\hline
Old stars \par t $>$ 1 Gyr&
Optical to near IR continuum&
Age, mass, formation epoch&
19 &
$^{}$ \par 10$^{11 -12}$
%$>10^{9}$
\\
\hline
Young stars \par t $<$0.5\,Gyr&
UV-optical&
Star formation rates, ages&
20,8 &
$^{}$ \par 10$^{9 - 10}$
\\
\cline{2-4}
 &
Ly$\alpha $&
Star formation rate&
20 &
 \\
\hline
\hline
Quasar (hidden or dormant)&
UV-optical \par polarisation \par broad lines&
Luminosity&
21,22&
 \\
\hline
Supermassive black hole&
Extended radio, \par
Quasar &
Formation, evolution &
23,24&
$^{}$ $\sim$10$^{9}$
 \\
\hline
\end{tabular}
\end{center}
\end{table}

\begin{figure}[ht!]
\centering
%%% Use the relevant command to insert your figure file.
%%% For example, with the graphicx package use
\vskip 8.0cm
\includegraphics{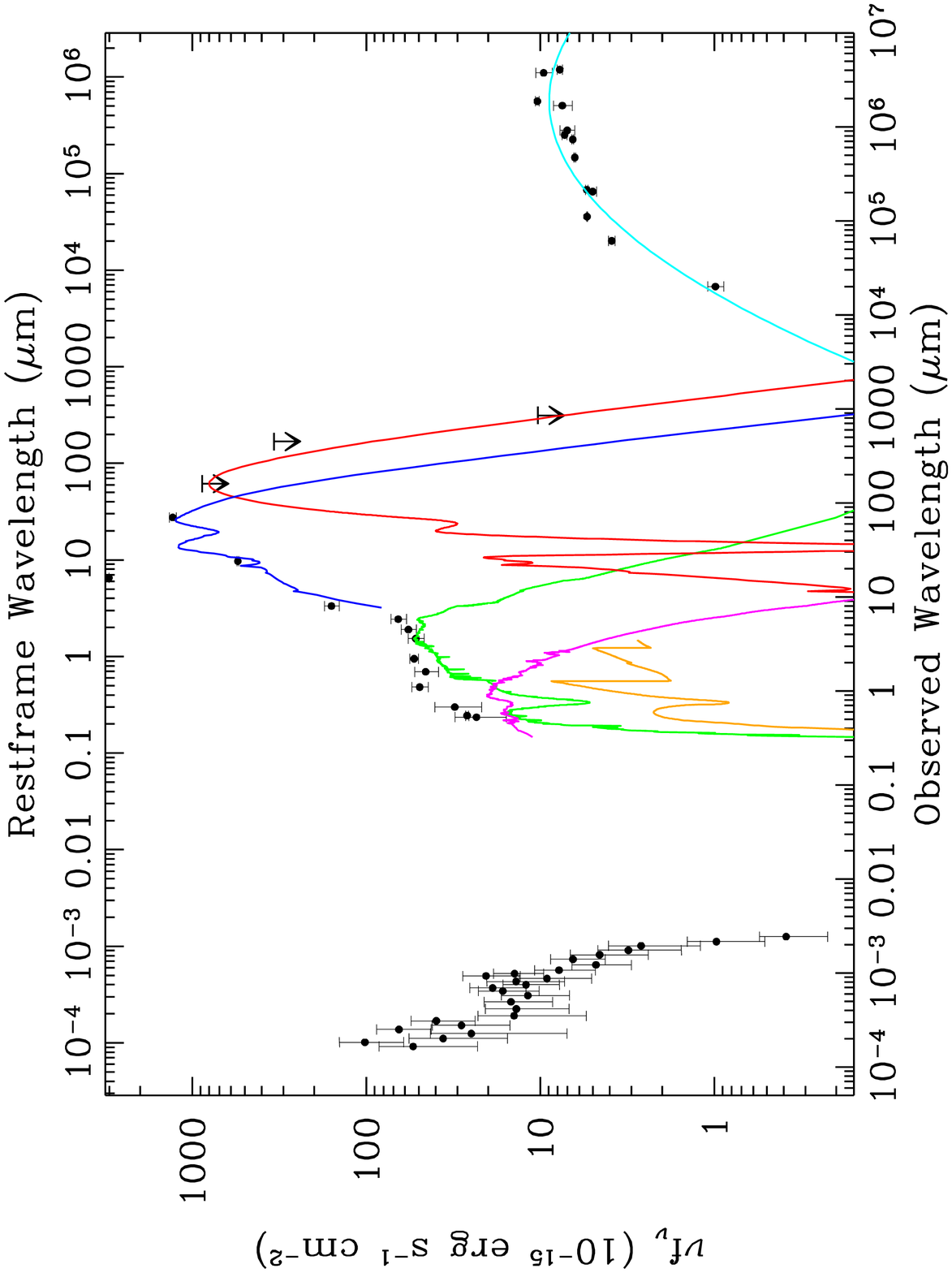}
%\plotone{width=0.75\textwidth,angle=-90]{2356_sed2.eps}
%%% figure caption is below the figure
\caption{Spectral energy distribution (SED) of the continuum emission from the HzRG 4C23.56 at z = 2.5, illustrating 
the contributions from the various constituents. [From De Breuck et al. in preparation]. Colored lines show the decomposition of the SED into individual components, under many assumptions. 
Cyan = radio synchrotron; Black = Absorbed nonthermal X-ray AGN; Yellow = nebular continuum; Blue = AGN-heated thermal dust emission;  Red = Starburst-heated dust emission; Green = Stars; Magenta = scattered quasar. The addition of the overlapping modeled components fits the SED well. 
%Grey = sum of the previous three components (stars,scattered quasar and nebular).
}
\label{fig: sed}       % Give a unique label
\end{figure}

Because they are highly luminous and (unlike quasars) spatially
resolvable from the ground, most components of HzRGs provide important
diagnostic information about the spatial distributions of processes
within HzRGs and their environment. The fact that the different
constituents are present in the same objects and that the {\bf {\it
interrelationships and interactions between them}}\ can be studied
make distant radio galaxies unique laboratories for probing massive
galaxy and cluster formation in the early Universe.

\vskip 0.2in
\centerline{\bf 2. The Spiderweb Galaxy - a case study}
\label{spider}

The Spiderweb Galaxy (MRC 1138--262), at a redshift of $z = 2.2$, is one
of the most intensively studied HzRGs. This object
provides a useful case study for illuminating several important
physical processes that may occur generally in the evolution of the
most massive galaxies. Because the Spiderweb Galaxy is (i) relatively
close-by, (ii) one of the brightest known HzRGs and (iii) the HzRG
with the deepest HST optical image, it is an important laboratory for
testing simulations of forming massive galaxies at the centers of
galaxy clusters.

This large galaxy has several of the properties expected for the
progenitor of a dominant cluster galaxy. The IR luminosity corresponds
to a stellar mass of $\sim $10$^{12 }$M$_{\odot}$,
implying that MRC 1138--262 is one of the most massive galaxies known
at z $>$ 2.  The host galaxy is surrounded by a giant Ly$\alpha $ halo
and embedded in dense hot ionized gas
with an ordered magnetic field causing rotation measures up to $\sim
6000$ rad m$^{-2}$. The radio galaxy is associated with
a 3 Mpc-sized structure of galaxies, of estimated mass $>2\times10^{14}$ 
M$_{\odot}$.  This type of structure is the presumed antecedent of 
present-day rich clusters.

\begin{figure}[t]
\plottwo{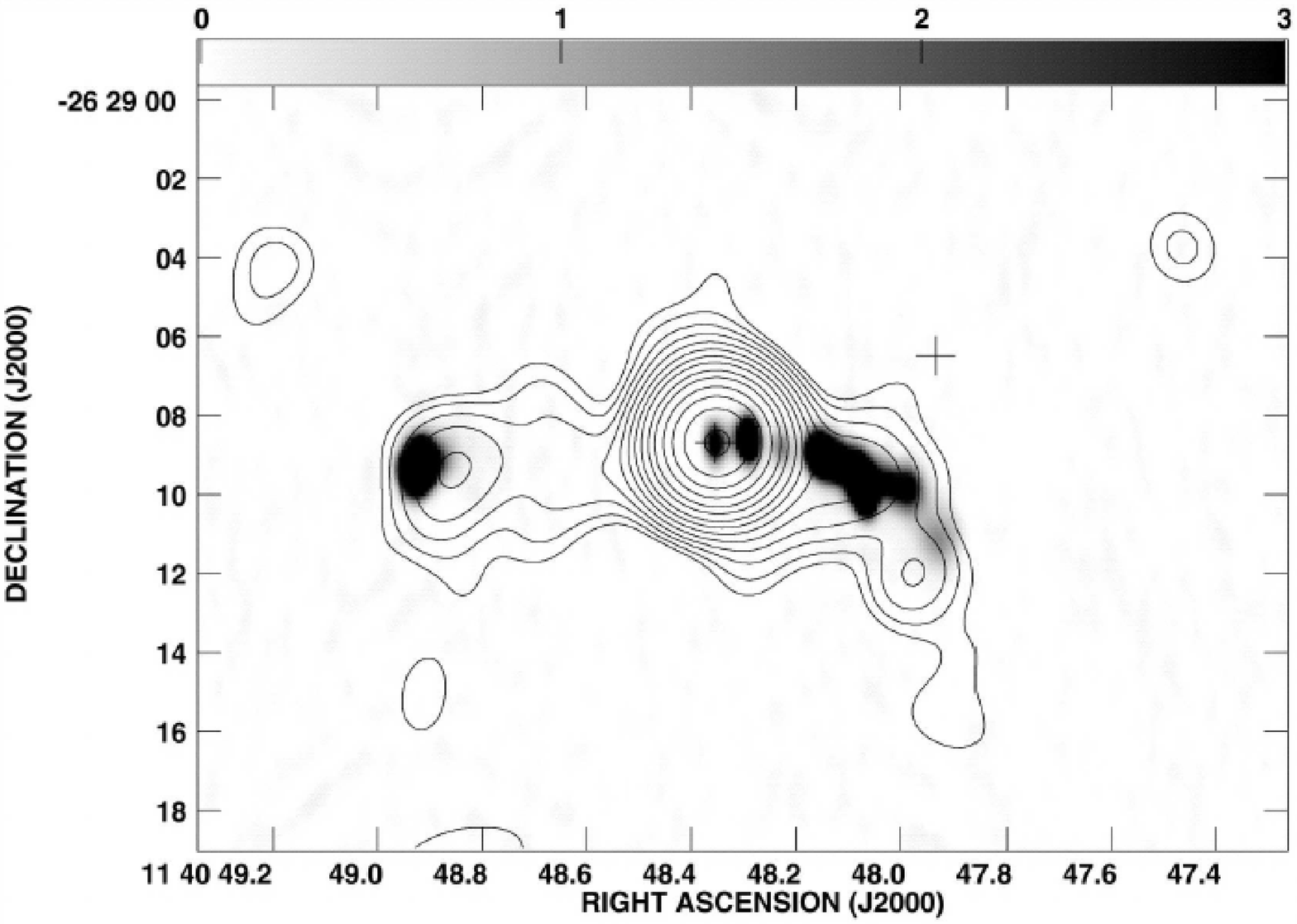}{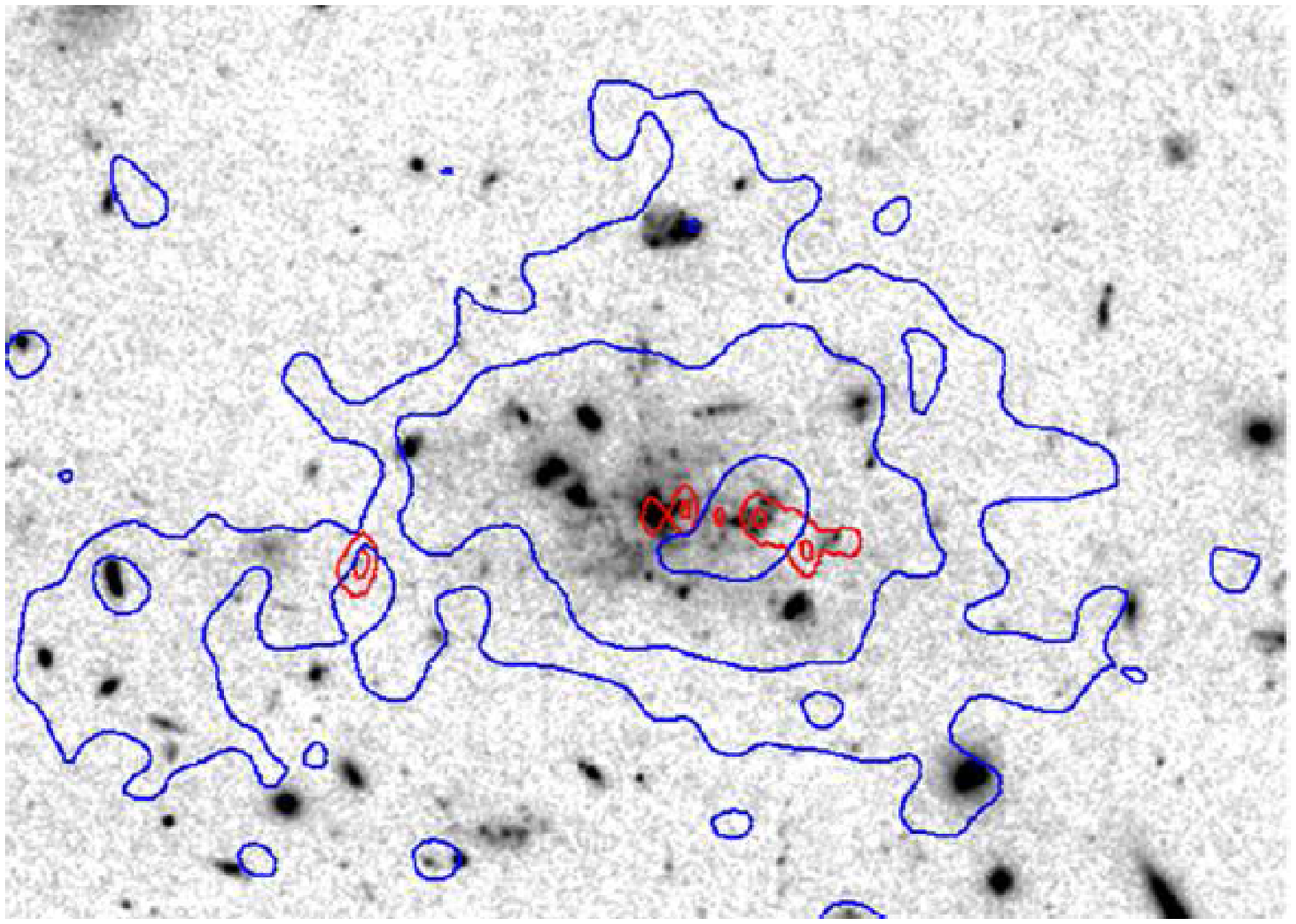}
\caption{{\bf Left:} X-ray emission from the Spiderweb Galaxy, PKS 1138-262 at z = 2.2, observed with the Chandra X-ray Telescope (2). X-ray contours are superimposed on a VLA gray scale representation of the 5 GHz radio continuum emission at 0.5'' resolution. The cross marks the position of the radio galaxy nucleus. Note that the X-ray and radio emission are aligned with each other.
{\bf Right:} The Spiderweb Galaxy. Deep Hubble image of the core of the MRC 1138-262
protocluster at z = 2.2 obtained with the Advanced Camera for Surveys. [From (8).
Superimposed on the HST image are contours of Ly$\alpha $ (blue, resolution
$\sim $1'') obtained with ESO's very Large Telescope (VLT), delineating the
gaseous nebula and radio 8 GHz contours (red, resolution 0.3'') obtained with
NRAO's VLA, delineating the non-thermal radio emission. The gaseous nebula
extends for $>$200 kpc and is comparable in size with the envelopes of cD
galaxies in the local Universe.\label{fig: spider}}
\end{figure}

The beautiful \textit{ACS} Hubble image of the Spiderweb Galaxy is
shown in Figure \ref{fig: spider}.  This figure illustrates the structures of
the radio, warm gas and stellar components in a relatively nearby
HzRG.  It also provides dramatic evidence that tens of satellite galaxies
were merging into a massive galaxy, $\sim$10 Gyr ago. The
morphological complexity and clumpiness agrees qualitatively with
predictions of hierarchical galaxy formation models, and illustrates this
process in unprecedented detail.  Ly$\alpha$ spectroscopy shows
relative velocities of several hundred km s$^{-1}$, implying that the
satellite galaxies (''flies") will traverse the 100 kpc extent of the
Spiderweb many times in the interval between z $\sim $ 2.2 and z $\sim
$ 0, consistent with the merger scenario.

An intriguing aspect of the Spiderweb Galaxy is the presence of faint
diffuse emission between the satellite galaxies. Approximately $50\%$
of the ultraviolet light from the Spiderweb Galaxy is in diffuse
''intergalactic" light, extending over a $\sim$60 kpc diameter
halo. The luminosity in diffuse light implies that the emission is
dominated by young stars with a star formation rate of $>$80 $M_\odot$
yr$^{-1}$. Under reasonable assumptions, the diffuse emission seen in
the Spiderweb Galaxy could evolve into the CD envelopes seen in many
dominant cluster galaxies at low redshifts.

The total mass of all the flies in the Spiderweb, derived from their
UV luminosities (assuming 1Gyr starbursts), is less than a tenth of
the mass of the whole galaxy obtained from its IR luminosity. Because 
the UV emission is produced by ongoing star formation and the IR 
emission by old stars, this implies that most of the galaxy mass may 
already have been assembled by z $\sim $ 2.2, consistent with downsizing 
scenarios.

Merging, downsizing and feedback are all likely to be occurring
simultaneously in the Spiderweb Galaxy. Merging is a plausible fueling
source for the nuclear supermassive black holes that produce the radio
sources. Pressure from these radio sources is sufficient to expel a
large fraction of gas from the galaxies, thereby
quenching star formation.  Because radio lifetimes are
relatively short (few $\times $10$^7$yr), all massive ellipticals may
have gone through a similar short but crucial radio-loud phase during
their evolution.

An unexpected feature of the HST image is that there is a significant
excess of faint satellite galaxies with linear structures
(8).  These galaxies (linear ''flies") have similar
morphologies (e.g. chains and tadpoles) to the linear galaxies that
dominate resolved faint galaxies (i$_{775 }>$ 24) in the Hubble Ultra
Deep Field (UDF) (4, 10). Although linear galaxies must be
an important constituent of the earliest galaxy population, their
nature is poorly understood.  Their presence in a merging system is
relevant for theories of their formation.  In the Spiderweb Galaxy the
motions of the flies with velocities of several hundred km s$^{-1}$
through the dense gaseous halo, perturbed by superwinds from the
nucleus (1, 13) and the radio jet, and/or more frequent
interaction of the flies with each other would result in shocks. The
shocks would then lead to Jeans-unstable clouds, enhanced star
formation along the direction of motion and to chain and tadpole
morphologies (11, 8).

\vskip 0.2in
\centerline{\bf 3. Unanswered questions}

There are still many aspects of HzRGs that are not understood, including:

%\begin {enumerate}
\begin{itemize}\addtolength{\itemsep}{-0.5\baselineskip}

\item What is the particle acceleration mechanism in the relativistic
plasma and why do HzRGs have much steeper radio spectra than nearby
radio sources>

\item What caused the most luminous radio galaxies and quasars to
become virtually extinct between z $\sim$ 2 and the present?

\item What is the nature, extent and kinematics of the hot gas, one of
the most massive and least studied constituents of HzRGs?

\item What is the origin of the Ly$\alpha$ halo that appears to be
falling into the HzRG and how is this warm gas (filling factor $\sim$
10$^{-5}$) distributed with respect to the hotter and colder gas
(filling factors $\sim$ 1)?

\item What are the processes by which the radio jets interact
with the gas and trigger starburst, and how important is jet-induced
star formation in the early Universe?

\item What are the temperature and densities and composition of the
  molecular gas and dust? What are their spatial distributions
  and what does this imply for the star formation histories?

\item What is the relative importance of AGN and starburst heating in
the mid-IR emission?

\item What effect does galaxy merging have on star formation and what
physical effects are responsible for downsizing?

\item What effect does feedback between the AGN and the galaxy have on
the evolution of HzRGs and the general evolution of massive galaxies?

\item What is the detailed mechanism by which the SMBHs produce
quasars and jets?

\item How and why do the AGN/quasars vary on long time-scales?

\item How is the SMBH built up and what role does merging play in this
evolution?

\item What is the size distribution of radio-selected protoclusters
and what is is the topology of the cosmic web in the neighbourhood of
the HzRGs?

\item How do the various populations and constituents of distant
protoclusters evolve and eventually become virialised clusters?

\item What is the radio luminosity function of galaxies in
radio-selected protoclusters?  Is the radio emission from the
protocluster galaxies an important contributor to forming the radio
halos, seen at the centres of many nearby rich clusters?
What is the strength and configuration of the protocluster magnetic
fields?

%\end{enumerate}
\end{itemize}

\vskip 0.15in
\centerline{\bf 4. A promising future}

There are good prospects for making progress on the many unknown aspects
of HzRGs during the next decade. Since its inception, the study of radio 
galaxies has been
observationally driven.  Several forefront astronomical facilities are
now being constructed or planned that will give new insights into the
nature of HzRGs and their environments. 

First, with a combination of sensitivity and spatial resolution, the
new low-frequency radio arrays, LOFAR and the LWA, will open up the
frequency window below $\sim$ 50 MHz for HzRG studies. The LWA will
survey the sky to unprecedented depth at low-frequencies and will
therefore be sensitive to the relatively rare radio sources that have
extremely steep spectra. Because of the $\alpha$ vs z relation, the LWA
is likely to detect HzRGs at z $\sim$ 8, if they exist.  Studies of
detailed low-frequency spectra and their spatial variations will
provide new information about the mechanism responsible for the
$\alpha$ vs z relation.  Presently combinations of the new radio
surveys with planned new deep optical and infrared wide-field surveys,
such as PAN-STARRS (6) and those with the VST and VISTA
will be used to identify HzRGs and provide photometric
data.

Another task for sensitive radio arrays, such as LWA, LOFAR, the EVLA
and eventually the Square Kilometre Array (SKA), will be 
to survey the radio emission of galaxies in protoclusters. The new
arrays will study radio emission produced by relativistic jets and be
able to detect and investigate radio emission from the brightest star
forming galaxies.

Secondly, ALMA and the EVLA, with their unprecedented
sensitivities and resolutions at millimetre and sub-millimetre
wavelengths, will revolutionize the study of molecular gas and
dust. Several different CO transitions can be observed, allowing
entire ``CO ladders'' to be constructed and the density and
temperature structure of the molecular gas to be unraveled. Fainter
molecular lines can be used to trace even denser gas than that studied
until now. Important information about the dust composition and the
gas to dust ratios is likely to be obtained.

ALMA's sensitivity at millimetre wavelengths should also facilitate
observations of the atomic CI Carbon lines in HzRGs. This would
provide an important constraint on the global metalicity of the
gas. The fine-structure line of C$^+$ at $\lambda_{\rm
rest}$=157.74$\mu$m line is one of the main cooling lines in nearby
galaxies, and has now also been detected in several of the most
distant quasars known (12).

Thirdly, we can expect considerable progress in disentangling the
detailed evolutionary history both of HzRGs and of radio-selected
protoclusters. This evolutionary detective work will be pursued by
combining spectroscopic data from the next generation of spectrographs
on 8m-class telescopes with imaging results from the new camera, WFC3
(7), on the Hubble Space Telescope. For example, the
detection of supernovae in z $\sim$ 2 protoclusters will become
possible.  On a longer timescale, tracing the detailed history of the
formation and evolution of HzRGs and the surrounding protoclusters
will be helped enormously with the advent of 30m -class ground-based
telescopes in the optical and near-infrared and the James Webb Space
Telescope (JWST) in the near and mid-infrared.

Fourthly, the next generation X-ray telescope, such as XEUS or
Constellation-X, will make observations of X-rays from HzRGs an
important tool for studying galaxy formation. It will have sufficient
sensitivity to perform spectroscopic studies of hot gas in HzRGs.  The
0.3 - 10 keV X-ray band contains the inner (K-shell) lines for all of
the abundant metals from carbon to zinc as well as many L-shell
lines. These atomic transitions provide important new plasma
diagnostics of the HzRG hot gas.

Fifthly, and perhaps most exciting, the potential discovery of HzRGs
with z $>$ 6 could open up a unique new window for studying the very
early Universe during the "Epoch of Reionisation". Recent
observational constraints suggest that cosmic reionization may have
taken place between z $\sim$ 11 and z $\sim$ 6. The existence of HzRGs
within the near edge of cosmic reionization could be used as sensitive
probes of intermediate- to small-scale structures in the neutral IGM,
through redshifted HI absorption observations (3, 5),
complementary to the very large scale that can be studied in HI
emission.

\end{document}